\documentclass[10pt,letterpaper]{article}
\usepackage{opex3,amsmath,amssymb}
\usepackage{xcolor}

\newcommand{\bc}{\begin{cases}\begin{aligned}}
\newcommand{\ec}{\end{aligned}\end{cases}}
\newcommand{\eq}{\begin{equation}}
\newcommand{\fine}{\end{equation}}
\newcommand{\beq}{\begin{equation}}
\newcommand{\eeq}{\end{equation}}

\newcommand{\uno}{\leavevmode\hbox{\small1\normalsize\kern-.33em1}}
\newcommand{\xx}{{\bf x}}

\newcommand{\casi}{\begin{cases}\begin{aligned}}
\newcommand{\casiend}{\end{aligned}\end{cases}}

\newcommand{\re}{\Re {\rm  e}}

\newcommand{\normCIB}{{\mathcal N}}
\newcommand{\CiB}{{\rm CiB}_{p,\ell}^{(\xi,q_0)}}

\renewcommand{\eqref}[1]{(\ref{#1})}

\begin{document}

\title{Birth and evolution of an optical vortex}

\author{Giuseppe Vallone,$^{1,*}$ Anna Sponselli,$^2$ Vincenzo D'Ambrosio,$^{3,5}$
Lorenzo Marrucci,$^4$ Fabio Sciarrino$^3$ and Paolo Villoresi$^1$}

\address{
$^1$Dipartimento di Ingegneria dell'Informazione, Universit\`a di Padova, I-35131 Padova, Italy
\\
$^2$Dipartimento di Fisica e Astronomia, Universit\`a di Padova, I-35131 Padova, Italy
\\
$^3$Dipartimento di Fisica, Sapienza Universit\`a di Roma, I-00185 Roma, Italy
\\
$^4$Dipartimento di Fisica, Universit\`a di Napoli Federico II and CNR-ISASI, Napoli
\\
$^5$Present address: ICFO-Institut de Ciencies Fotoniques, Barcelona Institute of Science and Technology, 08860 Castelldefels, Spain}

\email{$^*$vallone@dei.unipd.it} 

\begin{abstract}
When a phase singularity is suddenly imprinted on the axis of an ordinary Gaussian beam, an optical vortex appears and starts to grow radially, by effect of diffraction. This radial growth and the subsequent evolution of the optical vortex under focusing or imaging can be well described in general within the recently introduced theory of circular beams, which generalize the hypergeometric-Gaussian beams and which obey novel kinds of ABCD rules. Here, we investigate experimentally these vortex propagation phenomena and test the validity of circular-beam theory. Moreover, we analyze the difference in radial structure between the newly generated optical vortex and the vortex obtained in the image plane, where perfect imaging would lead to complete closure of the vortex core.
\end{abstract}




\section{Introduction}
Research on the properties of optical vortices started with the seminal work of J. Nye and M. Berry~\cite{nye74prs}.
There, the optical vortex was defined as
a line along which the phase of the electromagnetic field is indeterminate, namely a line on which the intensity is zero. 
Beams carrying a defined photon value of the orbital angular momentum (OAM) are characterized by an optical vortex on the beam axis:
the phase integration around the vortex divided by $2\pi$
gives an integer value, $\ell$, corresponding to the OAM content of the 
beam (per photon).

The observation that Laguerre-Gaussian (LG) modes have a well-defined OAM~\cite{alle92pra}
has prompted, in the last decades,  the investigation of the properties of beams carrying OAM 
and the possible techniques to generate, manipulate and analyze them.
{Nowadays, several devices are used to generate OAM beams: spiral phase plates~\cite{beij94opc},
diffractive elements such as holograms~\cite{hack92opl} or spiral Fresnel lenses, 
cylindrical lenses, spatial light modulators
and $q$-plates \cite{marr06prl,picc10apl,damb15sre} (see \cite{yao11aop} for a general review on the devices used to generate OAM)}. 

Each of them, applied to a plane wave (or  to a Gaussian beam), generates an optical vortex by 
multiplying the field by the phase profile $\exp(i\ell\phi)$, where $\phi$ is the angular coordinate along the plane transverse to the  propagation.
For ideally thin elements, this occurs with no other alterations of the beam field profile at the device output plane (sudden approximation).
The study of the dynamic of an optical vortex has important applications in many fields, such as long distance
free-space propagation~\cite{damb12nco,vall14prl} or microscopy with OAM beams.
{The first study on the generation of an optical vortex~\cite{sack98josaB} investigated the creation}
of filaments and the possibility of
having vortex sizes two orders of magnitude smaller than the overall beam size.
The propagation features of beams with OAM were studied in~\cite{sund05opl} and recently, a general theorem
on the divergence of vortex beam was demonstrated~\cite{vall16qph}.
In~\cite{kotl05josaA}, the far-field intensity distribution obtained with plane waves or Gaussian beams impinging on a
phase plate was derived, {while \cite{gbur16opt} theoretically studied the propagation of a Gaussian beam on which a 
fractional vortex plate is applied}.
In~\cite{kari07opl}, the Hypergeometric-Gaussian (HyGG) modes were introduced and it was shown that they can
be obtained by applying the singular phase factor to a Gaussian-parabolic transmittance profile.

However, the different studies lack a comprehensive model for the vortex propagation through generic systems.
Here we fill this gap by presenting an analytical model that  quantitatively describes
birth, evolution and 
closure of an optical vortex passing through an optical system that includes free space and two lenses.
We also experimentally verified the validity of this model. 
The main tool that we use is the recently introduced family of Circular beams (CiBs)~\cite{band08opl,vall15opl},
representing a very general solution of the paraxial wave equation with OAM.
The use of CiBs is  motivated by their simple transformation law under a generic ABCD optical system 
(see eq. \eqref{ABCDxi}).
 Since many well known beams carrying OAM are CiBs with particular values of the beam parameters,
 the ABCD law can be also easily applied to them. 
Then, in the present work we experimentally demonstrate that the CiBs represent 
 an optimal model for the study of the properties of optical vortexes
propagating through generic paraxial optical systems.

\begin{figure}[t]
\centering
\fbox{\includegraphics[width=8cm]{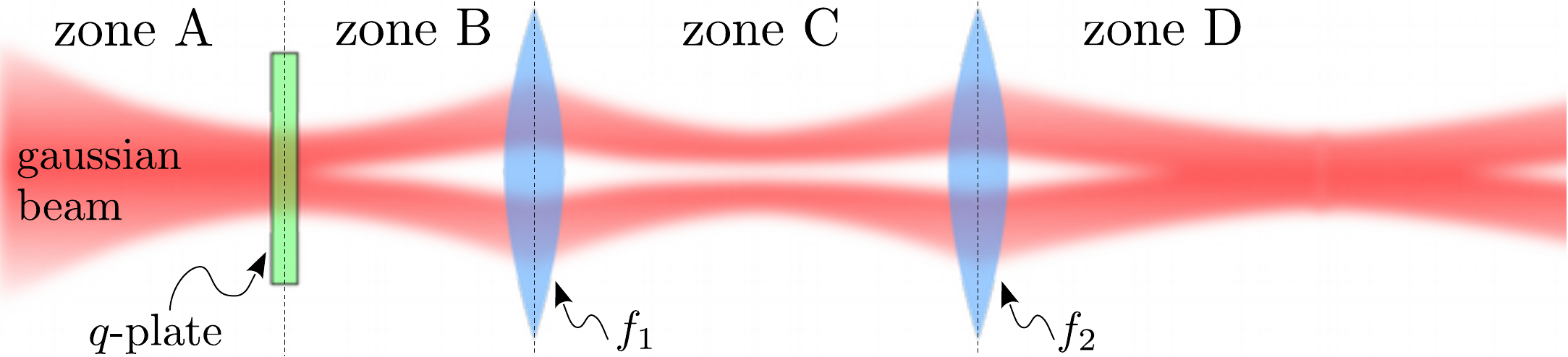}}
\caption{Experimental setup. A Gaussian beam is sent through a $q$-plate that is placed at the beam waist location
and generates the optical vortex. 
The obtained beam passes then through two lenses with focal lengths $f_1$ and  $f_2$. 
This defines three interesting propagation zones which will be analyzed: (B) newly generated vortex in free propagation; (C) focused vortex; (D) controlled imaging of the vortex source plane.}
\label{fig:setup}
\end{figure}

\section{Circular beams} 
The Circular beams, that will be used to model the birth and evolution of an optical vortex, are
here introduced.
We define the beam propagation direction as the $z$ axis and
we use polar coordinates $\xx=(r,\phi)$ in the plane transverse to the propagation. 
A generic CiB is determined by three complex parameters $\xi$, $q_0:=-d_0+iz_0$ and $p$ and one 
integer parameter $\ell\in\mathbb Z$.
In a given transverse plane (say, $z=0$), the 
monochromatic CiBs is defined as:
\beq
\label{cib}\CiB
=\normCIB
(1+\xi\frac{q^*_0}{q_0})^{\frac {p}2}
{(\frac{i\sqrt{k z_0}\,r}{q_0})}^{{|\ell|}}
{\rm G}(r)
\ _1F_1(-\frac p2,|\ell|+1;\frac{r^2}{\chi^2})
e^{i\ell\phi}\,.
\eeq
In the above equation 
${\rm G}(r)=i\sqrt{\frac{k z_0}{\pi}}e^{-\frac{ikr^2}{2q_0}}/q_0$ is the Gaussian beam,
$k$ is the wavevector, $\chi$ is defined by $\frac{1}{\chi^2}=
\frac{kz_0\xi}{q_0}\frac{1}{q_0+\xi q^*_0}$,
$_1F_1$ is the Hypergeometric function 
and $\normCIB=
[ |\ell|! _2F_1(-\frac p2,-\frac{p^*}{2},|\ell|+1,|\xi|^2)]^{-1/2}$ 
is a normalization factor depending on $p$, $|\ell|$ and $|\xi|$~\cite{vall15opl}. 

We now give a physical interpretation of the parameters characterizing the CiBs.
The first, $\xi$, is related to the ``shape'': specific values of $\xi$ identify some well-known beams.
For instance, 
the limit $\xi\rightarrow+\infty$ corresponds to the standard LG modes~\cite{sieg86lasers}.
CiBs with $|\xi|=1$ correspond to the generalized HyGG~\cite{kari07opl,vall15opl,kari08opl}:
in particular, the HyGG defined in \cite{kari07opl} or
 the HyGG-II defined in \cite{kari08opl} are obtained by setting $p\in\mathbb R$ and
$\xi=1$ or $\xi=-1$, respectively.
We will show below that $\xi$ is also related to the properties of the beam under the propagation through optical systems.
The parameter $q_0$ is related to the physical scale (similarly to the complex beam parameter of the Gaussian beam~\cite{sieg86lasers}):
its imaginary part is $z_0=kw^2_0/2>0$ with
$w_0$ the analog of the Gaussian ``beam waist'' while $d_0$ represents the location of the beam waist.
Finally, $p$ defines the radial index and $\ell$ corresponds to the carried OAM.

CiBs were originally defined~\cite{band08opl} in terms of $(q_0, q_1)$, with $q_1=\frac{q_0+\xi q^*_0}{1+\xi}$.
However, the formulation in terms of $(q_0, \xi)$ gives a clearer view
when the  propagation in free space and throughout a generic ABCD
system is studied.
In particular, as we will show, the parameter $|\xi|$ defines a ``class'' of beams
with similar features under the propagation through generic optical systems.
\begin{figure}[htb]
\centering
\begin{minipage}{10.2cm}
\includegraphics[width=10.2cm]{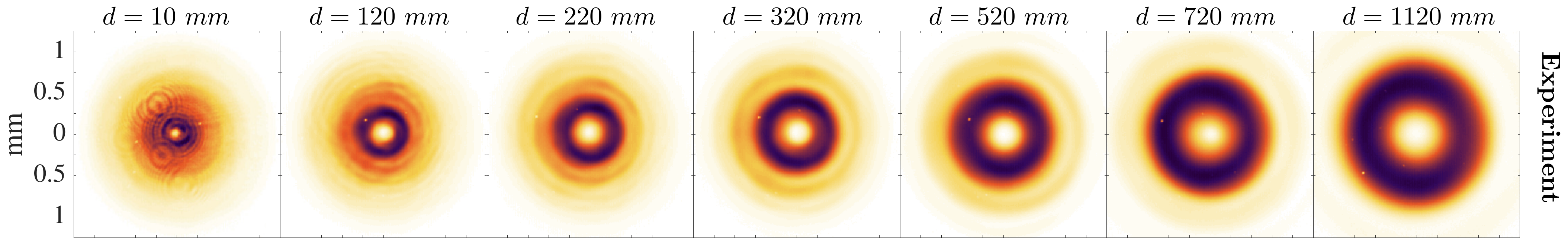}\vskip.02cm
\includegraphics[width=10.2cm]{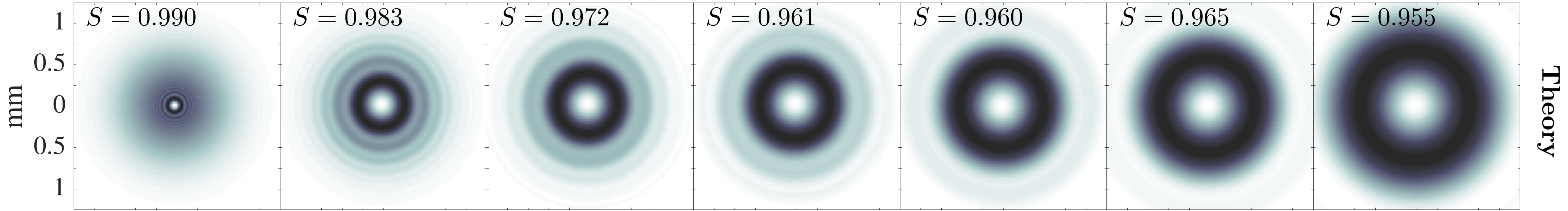}\vskip.02cm
\includegraphics[width=10.2cm]{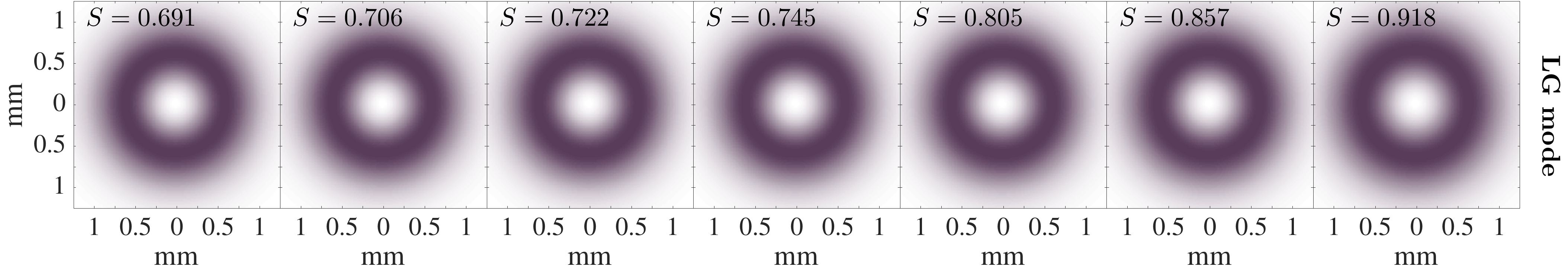}
\end{minipage}
\begin{minipage}{3cm}
\includegraphics[width=3cm]{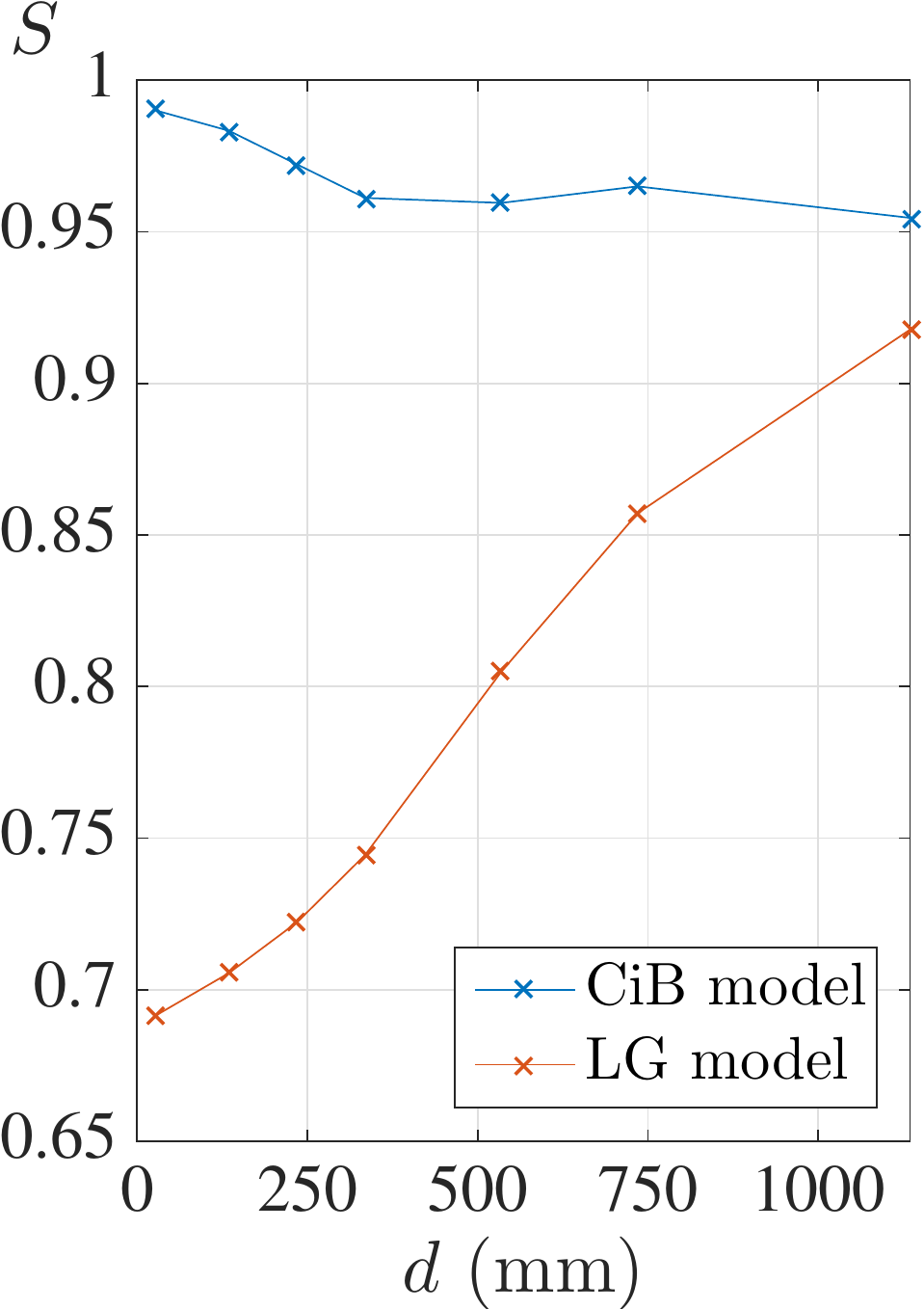}
\end{minipage}
\caption{``Birth'' of the optical vortex. In the first row we show the experimental intensity patterns obtained after
 the $q$-plate placed at the beam waist location
at various propagation distances $d$. The data should be compared with the theoretical CiB model 
shown in the second row. In the third row we show the corresponding Laguerre-Gauss mode with the same $z_0$
and beam waist located at $d=0$. 
The degree of agreement between the two models and the experiment 
is measured by the reported similarity values, $S$, given in each panel and shown in the right inset for different values of $d$.}
\label{fig:vortex}
\end{figure}
As derived in~\cite{band08opl}, the field resulting after a propagation along a distance $d$ can be
obtained from \eqref{cib}
by the transformation $q_j\rightarrow q_j+d$ with $j=0,1$.
Such law is generalized for generic ABCD optical system in terms of $(q_0,q_1)$ as
$q_j\rightarrow (A q_j+B)/(C q_j+D)$,
with $A,B,C,D\in \mathbb R$~\cite{band08opl}.
By using $(q_0,\xi)$ as beam parameters,
 the ABCD transformation can be rewritten as follows:
\beq
\label{ABCDxi}
q_0\rightarrow \frac{A q_0+B}{C q_0+D}
\,,\qquad
\xi\rightarrow \frac{C q^*_0+D}{C q_0+D}\xi\,.
\eeq
It is worth noticing that, by the above transformation, the absolute value of $\xi$ 
and the normalization $\normCIB$ are invariant.
Then, the parameter $|\xi|$ identifies an ``equivalence class'' of CiBs under generic ABCD optical transformations:
beams with different $|\xi|$'s cannot be obtained as input and output of any real ABCD system.
Moreover, as noticed in \cite{vall15opl}, the value of the parameter $|\xi|$ determines the constraints on $p$ to achieve square 
integrability. When $|\xi|<1$ the beam is square integrable $\forall p$; when $|\xi|=1$ it is required than $\re(p)>-1-|\ell|$;
when $|\xi|>1$ square integrability imposes $p/2\in \mathbb N$.
Finally, $\xi$ does not change in free-space propagation.
Then,  we believe that the description of CiBs in terms of $(q_0,\xi)$ offers a clearer view 
of their properties during generic optical transformations.

The  beams that can be easily experimentally generated by using phase plates or $q$-plates 
correspond to CiBs with $|\xi|=1$:
indeed, by applying the phase factor $e^{i\ell\phi}$ to a Gaussian beam with a given $z_0$ at a distance
$\Delta$ from its waist plane, the CiBs with $\xi=\frac{z_0-i\Delta }{z_0+i\Delta}$ and $p=-|\ell|$
is generated~\cite{kari07opl,vall15opl}.
After straightforward calculations from \eqref{cib}, such beams 
can be written as:
\beq
\label{kummer}
{\rm CiB}_{-|\ell|,\ell}^{(\xi,q_0)}=
\frac{\Gamma(|\ell|/2+1)}{|\ell|!}
{(\frac{-r^2}{ \xi\chi^2})}^{\frac{|\ell|}{2}}
{\rm G}(r)
\ _1F_1(\frac{|\ell|}2,|\ell|+1;\frac{r^2}{\chi^2})
e^{i\ell\phi}\,,
\qquad \text{when }|\xi|=1\,.
\eeq

The beam in \eqref{kummer}, that carries $\ell$ units of OAM, is quite different from one of the well known (and very much used)
LG modes. Indeed, the intensity of a LG$_{0,\ell}$ mode presents an intensity pattern that has always
an hole in the center, even at its waist location, while in the beam of eq. \eqref{kummer} the vortex 
is absent at a distance
$\Delta$ from its waist plane, but starts to grow during propagation.
 The goal of our experiment is then to study the evolution of a beam generated by applying 
the phase $e^{i\ell\phi}$ to a Gaussian beam and to compare it with the CiB model.

\section{The experiment}
The experimental setup is shown in Fig. \ref{fig:setup}. A circular polarized Gaussian beam impinges on a $q$-plate placed
on the beam waist. The optical vortex is then generated 
(the same results could be obtained with any device able to imprint the phase profile $e^{i\ell\phi}$).
The obtained beam propagates through an optical system composed of two lenses with focal
lengths $f_1$ and $f_2$ respectively. 
We measured the intensity pattern of the beam by a CCD camera across four different zones, labeled as A, B, C and D (Fig. 1),
 and compared the results with the   model
provided by the CiBs.
As a reference, we first measured the properties of the impinging Gaussian beam at wavelength $810~nm$ without $q$-plate.
 Taking $z=0$ the location of the beam waist, we measured
the beam intensity at seven different distances $d$, ranging from $10cm$ to $1120cm$.
From them we derived the beam waist $w_0=850\pm10~\mu m$,
corresponding to a Rayleigh distance $z_0=2.80\pm0.07~m$. 
To compare the theoretical and experimental intensity
patters we used the {\it similarity}, defined as 
$
S=(\sum_{x,y}I^{\rm th}_{x,y}I^{\rm exp}_{x,y})
/\sqrt{\sum_{x,y}(I^{\rm th}_{x,y})^2\times \sum_{x,y}(I^{\rm exp}_{x,y})^2}
$
where  $I^{\rm exp}_{x,y}$ and $I^{\rm th}_{x,y}$ are the experimental and theoretical intensities at point $(x,y)$ respectively.
In all the cases 
we measured $S>0.98$.
\begin{figure}[t]
\begin{minipage}{6.6cm}
\centering
\includegraphics[width=6.5cm]{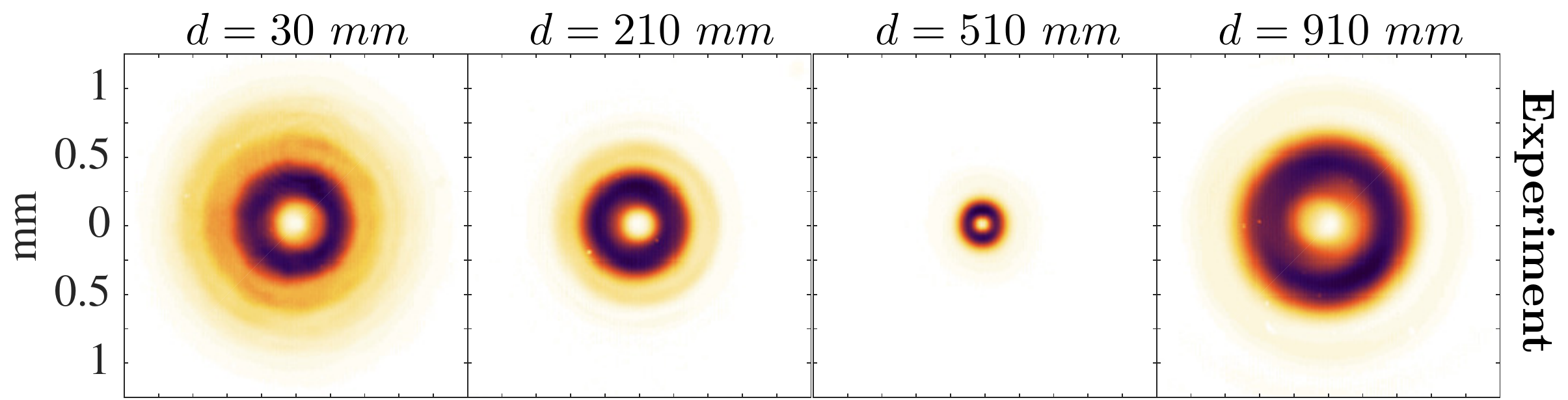}\vskip-.05cm
\includegraphics[width=6.5cm]{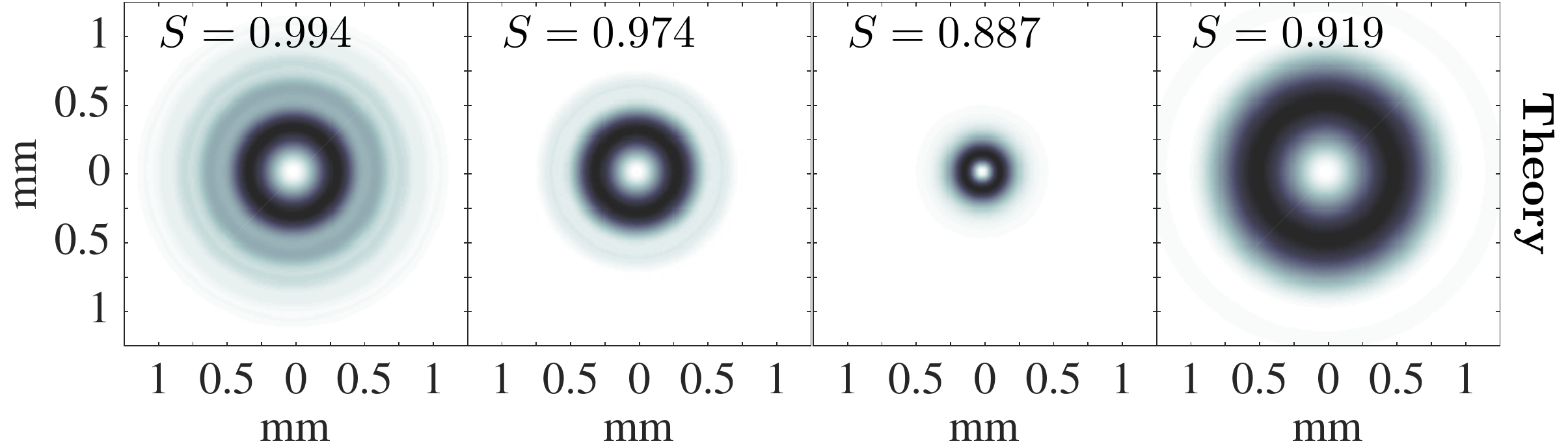}\vskip-.05cm
\end{minipage}
\begin{minipage}{6cm}
\centering\includegraphics[width=5cm]{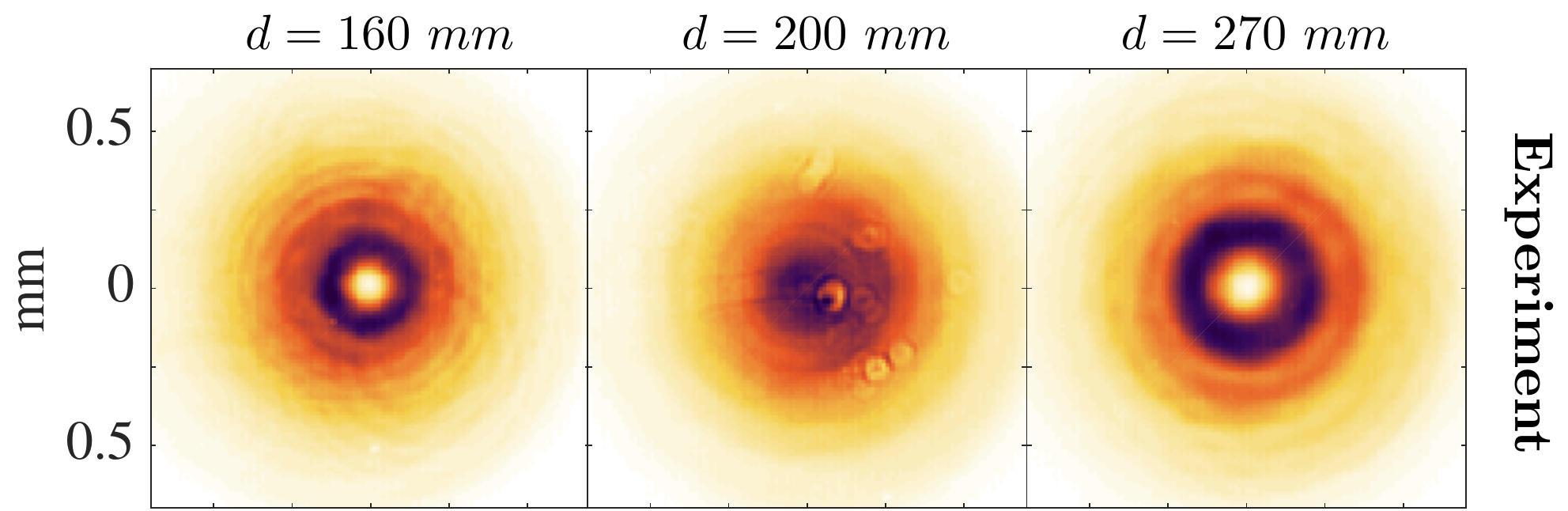}\vskip-.05cm
\includegraphics[width=5cm]{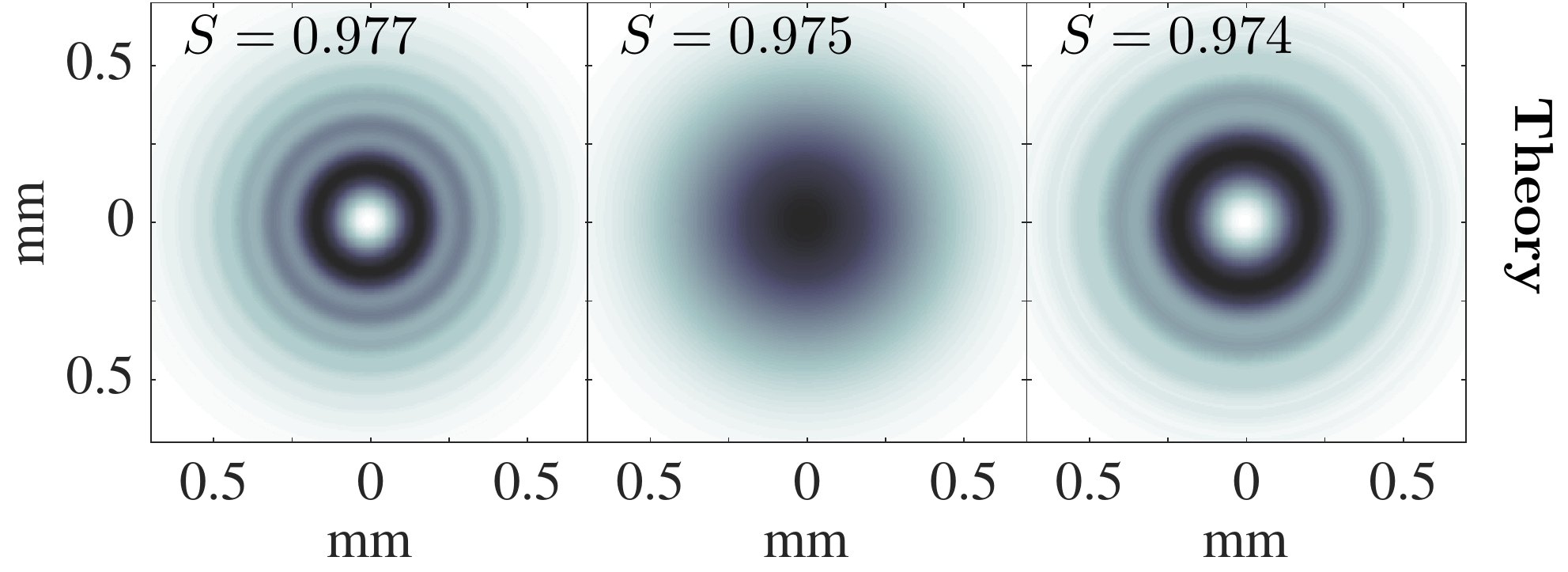}
\end{minipage}
\label{fig:vortex_lens}
\caption{{\bf Left}: Focusing of an optical vortex by a lens.
Experimental (upper row) and theoretical (lower row) intensity patterns obtained at various distances $d$ from the focusing lens.
{\bf Right}: Quasi-closure of the optical vortex
occurring when a real image of the vortex source is created by a lens system. 
Experimental (upper row) and theoretical (lower row) intensity patterns obtained at various distances $d$ from the second lens. 
The medium panel for $d = 200$ mm corresponds to the image plane.}
\label{fig:vortex_closure}
\end{figure}

We then placed a $q$-plate with $q=1/2$ at the beam waist location, to observe the ``birth'' of the optical vortex:
this corresponds to the ``zone B'' of Fig. \ref{fig:setup}. As predicted by
the CiBs model, the vortex, related to the phase singularity on the beam axis, starts growing after the $q$-plate.
After the  propagation 
 across a small distance, of the order of $z_0/10$,
the vortex diameter stabilizes to a size of the order of half beam diameter.
The ``birth'' of the optical vortex is shown in Fig. \ref{fig:vortex}, where we compare the experimental intensity patterns
with the theoretical model of \eqref{kummer} with $\ell=1$, $\xi=1$ and $q_0(d)=d+iz_0$. 
The good agreement between observed intensities and the predictions is proved by the similarities, always greater than 0.95:
we here stress that that no fitting parameter was employed in the theoretical model. Indeed, we used 
as Rayleigh distance the value $z_0=2.8~m$ obtained in the previous measurement.
The parameter $d$ reported in the top of Fig. \ref{fig:vortex} corresponds to the measured distance from the $q$-plate.
In Fig. \ref{fig:vortex} we also show the intensity of a LG mode with $p=0$, $\ell=1$ and the same $z_0=2.8~m$, to demonstrate
the differences between the experimental generated
beam, its theoretical model represented by the CiBs and the widely used LG approximation. 
While the beam resembles a Laguerre-Gauss mode in the far field, in the
near field (up to a distance of $\sim z_0/10$ from the $q$-plate) 
the LG approximation fails to correctly model the beam properties, as 
quantified by the low values of the similarities.
In particular, the ``birth'' of the optical vortex cannot be modeled by a LG mode, since, for such beams,
the vortex is present along all transverse planes during propagation, as shown for instance by the LG intensity
patterns at $d=10~mm$ in Fig. \ref{fig:vortex}.
The same conclusions were obtained in the theoretical analysis of \cite{beks08opc}: 
indeed, it is worth noticing that the subclass of CiB with $\xi=1$ and $p=-|\ell|$ corresponds to the ``Kummer beams'' defined in  \cite{beks08opc}.
However, the formulation in terms of CiBs, allows for a complete study of the propagation through generic optical systems,
as we now demonstrate.
Indeed, we also measured the beam
 evolution after a lens with focal length $f_1=510~mm$,
corresponding to the ``zone C'' of Fig. \ref{fig:setup}. The lens was placed at distance $d_1=150~mm$ from the $q$-plate plane
and now we indicate by $d$ the distance from the lens. By the ABCD law, the beam parameters
$(q_0(d), \xi_C)$ in zone $C$ can be calculated from $(iz_0,1)$
by the matrix
$M=\left(\begin{smallmatrix}
1-d/f_1 & \ \ d+d_1-dd_1/f_1
\\
-1/f_1 & 1-d_1/f_1
\end{smallmatrix}\right)$, obtaining 
\beq
q_0(d)=\frac{(d+d_1)f_1-dd_1+iz_0(f_1-d)}{f-d_1-iz_0}\,,\qquad \xi_C=\frac{f_1-d_1+iz_0}{f_1-d_1-iz_0}\,.
\eeq
In zone $C$, as expected, $\xi_C$ does not depend on $d$.
We show the comparison between the measured and calculated intensity patterns 
obtained at different distances $d$ from the lens in the left panel of Fig. \ref{fig:vortex_lens}.
The theoretical intensity patterns were evaluated by inserting in the CiBs of \eqref{kummer} the above values of $q_0(d)$ and $\xi(d)$.
Four of the eight measured patterns and the corresponding similarities $S$ are shown
(the lowest of the eight measured similarity is $0.885$): they demonstrate that CiB correctly models the propagation
of such beam. We notice that the chosen position of the lens is such that no real imaging of the $q$-plate takes place in zone C.

We finally observed the beam in the ``zone D'',
that is after a second lens leading to the formation of a real image of the $q$-plate vortex source. 
In this case we used $f_1=300~mm$ and $f_2=200~mm$, with
the lens $f_1$ placed at distance $d_1=f_1$ from the $q$-plate and the lens $f_2$ placed at distance $f_1+f_2$ from the
first lens. We now indicate by $d$ the distance after the lens $f_2$. The ABCD matrix in this case is given by
$M=\left(\begin{smallmatrix}
-f_2/f_1 & \ f_1-df_1/f_2
\\
0  &		-f_1/f_2
\end{smallmatrix}\right)
$
corresponding to $q_0(d)=d-f_1+iz_0f_2^2/f_1^2$ and $\xi_D=1$.
In Fig. \ref{fig:vortex_lens} (right) we show three of the seven measured patterns and the corresponding similarities $S$ 
(now the lowest measured similarity was $0.928$).
It is worth noticing the almost complete closure of the vortex: at the plane $d=200~mm$ the vortex at the center almost completely disappears,
even if the OAM content is still non-vanishing.
The disappearance is actually not complete: a more accurate analysis (to be reported elsewhere) shows that the vortex radius reduces to a minimum value that depends on the numerical aperture of the optical imaging system, according to the standard resolution limits imposed by wave theory. However, this minimum vortex size can easily be orders of magnitude smaller than the beam size in the same plane. 
We further notice that an approximate theory based on LG vortex beams does not describe properly this vortex imaging phenomenon.

\section{Conclusions}
We have studied the generation and the propagation of an optical vortex created by 
 superimposing an azimuthal phase pattern imprinted by a $q$-plate on a Gaussian beam. 
The application of such a phase mask is the principle 
on which all current approaches to generate and measure OAM eigenstates (spiral phase plates, fork holograms, $q$-plates) are based.
By using an optical system with two lenses, we have experimentally verified for the first time the recently 
introduced ABCD law for Circular beams~\cite{band08opl}.
Our results demonstrate that the CiBs are very useful to  analytically model
 the propagation through a generic optical system of OAM beams.
We stress that many well known beams carrying OAM are included in the CiB family 
with particular values of the beam parameters.
The accuracy of the $q$-plate in the generation of the 
singular phase profile $\exp(i\ell\phi)$ to the beam was essential to generate a 
 highly stigmatic beam
that perfectly matches the
theoretical predictions.

\section*{Acknowledgments}
Our work was supported by the 
Progetto di Ateneo PRAT 2013 (CPDA138592) of the University of Padova
and by the ERC-Starting Grant 3D-QUEST (3D-Quantum Integrated Optical Simulation; grant agreement no. 307783): 
http://www.3dquest.eu. 
LM acknowledges support of the ERC-Advanced grant PHOSPhOR 
(grant agreement No. 694683)

\end{document}